\documentclass[aps,prb,twocolumn,showpacs]{revtex4}
\usepackage{graphicx}

\begin{document}

\title{Thermal fluctuations and phase diagrams of the phase field crystal model with pinning}

\author{J.A.P. Ramos $^{1,2}$, E. Granato $^{2,4}$, C.V. Achim $^3$, S.C. Ying $^4$,
K.R. Elder $^5$, and T. Ala-Nissila $^{3,4}$}

\address{$^1$ Departamento de Ci\^encias Exatas, Universidade
Estadual do Sudoeste da Bahia,
45000-000 Vit\'oria da Conquista,
BA,Brasil}
\address{$^2$Laborat\'orio Associado de Sensores e Materiais,
Instituto Nacional de Pesquisas Espaciais,12245-970 S\~ao Jos\'e dos Campos, SP, Brazil}
\address{$^3$Department of Engineering Physics, P.O. Box 1100, Helsinki University of Technology,
FIN-02015 TKK, Espoo, Finland}
\address{$^4$Department of Physics, P.O. Box 1843, Brown University,
Providence, RI 02912-1843, USA}
\address{$^5$Department of Physics, Oakland University, Rochester,
Michigan 48309-4487, USA}


\begin{abstract}
We study the influence of thermal fluctuations in the phase diagram
of a recently introduced two-dimensional phase field crystal model
with an external pinning potential. The model provides a continuum
description of pinned lattice systems allowing for both elastic
deformations and topological defects. We introduce a non-conserved
version of the model and determine the ground-state phase diagram as
a function of lattice mismatch and strength of the pinning
potential. Monte Carlo simulations are used to determine the phase
diagram as a function of temperature near commensurate phases. The
results show a rich phase diagram with commensurate, incommensurate
and liquid-like phases with a topology strongly dependent on the
type of ordered structure. A finite-size scaling analysis of the
melting transition for the $c(2 \times 2)$ commensurate phase shows
that the thermal correlation length exponent $\nu$ and specific heat
behavior are consistent with the Ising universality class as
expected from analytical arguments.
\end{abstract}


\maketitle

\section{Introduction}

Many two-dimensional (2D) lattice systems on periodic potentials
form different ordered structures at low temperatures which may be
unstable against thermal fluctuations and the mismatch between the
competing periodicities. Adsorbed layers on crystal surfaces
\cite{patry,persson,patry07}, vortex lattices in 2D superconductors
\cite{martin97}  and colloidal crystals \cite{lin00,ling08} in a
periodic substrate are important examples of current interest. In
numerical simulations, pure elastic models or particle models are
often employed. However, there are important limitations in these
approaches. In the former, plastic deformations such as dislocations
are not taken into account and in the latter, the accessible time
and spatial scales may be very limited. These limitations are
particularly important in atomistic simulations of crystalline
materials described by microscopic models with complicated
interactions, where the time scale is set by the vibrational
frequency and consequently relatively small system sizes are often
used in numerical simulations.

Recently, a phase field crystal (PFC) model for pinned lattice
systems was introduced \cite{achim06} that allows for both elastic
and plastic deformations of the lattice. The model describes the
lattice system as a continuous  density field in an external pinning
potential. In this formulation a free energy functional is
introduced which depends on the phase field $\psi(\vec r)$
corresponding to the particle density. In absence of a pinning
potential, the free energy is minimized when $\psi$ is spatially
periodic \cite{elder02} forming a hexagonal solid phase in 2D. By
incorporating phenomena on short length scales the model naturally
includes elastic and plastic deformations and it should be
computationally more efficient than particle simulations such as
standard molecular dynamics \cite{elder02,elder04}.

In a previous work \cite{achim06}, the phase diagram of the PFC
model in a periodic pinning potential with square symmetry was
determined as a function of the pinning strength and lattice
mismatch, in the absence of thermal fluctuations. A numerical
minimization procedure was used to determine the different ordered
phases and the phase transitions, based on the corresponding
dynamical equation of motion for a conserved field. In the present
work, we consider the influence of thermal fluctuations and lattice
mismatch in the PFC model with pinning. We first introduce a
non-conserved version of the model which is more efficient for
equilibrium simulations, such that a constant chemical potential is
included in the free energy to fix the average density. With this
new version, we determine the ground-state phase diagram as a
function of lattice mismatch and strength of pinning potential using
the corresponding non-conserved dynamical equations of motion. Monte
Carlo (MC) simulations are then used to determine the phase diagram
as a function of temperature and mismatch near commensurate phases.
The results reveal that the model has a rich phase diagram, which
includes commensurate, incommensurate and liquid-like phases with a
topology dependent on the type of ordered structure. In particular,
a finite-size scaling analysis of the melting transition for the
$c(2 \times 2)$ commensurate phase shows that the thermal
correlation length exponent $\nu$ and specific heat behavior are
consistent with the Ising universality class, which is expected from
analytical arguments based on symmetry considerations and Landau
free-energy expansion.

\section{Model}

The {\it mean field} free-energy functional of the PFC model with an
external pinning potential \cite{achim06} can be written in
dimensionless units as
\begin{equation}
F = F_0 \int d \vec x \{  \frac{1}{2} \psi[r+(1+\nabla^2)^2] \psi +
\frac{\psi^4}{4} + V \psi \}, \label{cpfcp}
\end{equation}
where $\psi(\vec x)$ is a continuous field representing the local
number density of the particles and $V(\vec x)$ represents the
external pinning potential. The overall constant $F_0$ sets the
energy scale and $r$ is a parameter. In Eq. (\ref{cpfcp}),
$\psi(\vec x)$ is a conserved field and the minima of the free
energy depends on the average value of $\psi(\vec x)$.  To remove
this conservation constraint on $\psi(\vec x)$, we introduce here an
additional linear term to the free energy containing the chemical
potential $\mu$, which controls the average value of $\psi(\vec x)$.
This leads to a  model {\it mean field} free-energy functional
\begin{equation}
F = F_0 \int d\vec x \{  \frac{1}{2} \psi[r+(1+\nabla^2)^2] \psi +
\frac{\psi^4}{4} + V \psi -\mu \psi \}. \label{pfcp}
\end{equation}
This modification allows the use of a MC algorithm with non-conserved
dynamics as described in Sec. 3, which is more efficient for
equilibrium simulations.

In the absence of the pinning potential, $V(\vec x)=0$, the free
energy of Eq. (\ref{pfcp}) can be  minimized by a configuration of
the field $\psi(\vec x)$ forming a hexagonal pattern of peaks with
wavector of magnitude $k_0=1$, when the values of the parameters $r$
and $\mu$ are chosen appropriately. This structure of peaks can be
regarded as  a crystalline system and the  free energy functional of
Eq. (\ref{pfcp}) can then be used to describe both elastic and
plastic properties of such a lattice system within a mean field
description \cite{elder02,elder04,elder07} for temperatures $T$
below the mean field melting temperature $T_m$ by setting the
parameter $r \propto T- T_m$. However,  the free energy of Eq.
(\ref{pfcp}) do not provide all the information on the system at
finite temperatures and specially do not take into account strong
thermal fluctuations which may lead to the melting of the hexagonal
pattern.

To go beyond mean field theory and take thermal fluctuations into
account, we follow the usual procedure of introducing a
coarse-grained effective Hamiltonian \cite{chaikin} where the energy
for a particular configuration of the order parameter is given by
the corresponding mean field free-energy functional. Then instead of
considering just the minimum energy this procedure allows for
excitations according to a statistical weight provided by the free
energy functional. In the present case we take as the effective
Hamiltonian $H[\psi]\equiv F[\psi]$ with the corresponding partition
function
\begin{equation}
Z = \sum_{\{\psi(\vec x)\}} e^{- H[\psi]/k_BT}, \label{part}
\end{equation}
where $T$ is the temperature and the summation (functional
integration) is taken over all configurations of $\psi(\vec x)$.
Physical quantities are defined in the usual way as thermal averages
over $\psi$ configurations. The parameter $r$ contained in $H[\psi]$
is taken to be a temperature-independent constant for the region
below the mean field transition temperature $T_m$ in the absence of
the pinning potential.

We consider a pinning potential $V(x,y)$ with square symmetry
\begin{equation}
V(x,y)=V_0[\cos(k_s x)+ \cos(k_s y)],
\end{equation}
where $k_s = 2\pi/a_s$ is the  wave vector of the pinning potential,
which can represent, for example, and adsorbed layer on the $(100)$
face of an fcc crystal \cite{patry,patry07}. We define the lattice
misfit between the lattice system and pinning potential as
$\delta_{\rm m}=(k_0-k_s)/k_0$, where $k_0=1$ is the wave vector of
the hexagonal periodic pattern\cite{foot1} of $\psi(x,y)$ in the
absence of the pinning potential, corresponding to a free lattice
system. Thus, for $\delta_{\rm m} \gg 0 $, the lattice system is
under tensile strain while for $\delta_{\rm m} \ll 0$ it is under
compression.

\section{Monte Carlo simulation}

The free energy associated with the partition function of Eq.
(\ref{part}), which now takes into account thermal fluctuations can
not be obtained directly without approximations. Therefore, to
obtain the equilibration configurations and average quantities  it
is necessary to perform numerical simulations and for this purpose
we used the MC method. Thermal averages over configurations of
$\psi(x,y)$ were performed using MC simulations on a discrete
version of the effective Hamiltonian $H[\psi]$. The field
$\psi(x,y)$ and pinning potential $V(x,y)$ are defined on a square
space grid, $x = idx$, $y=jdy$ ($i,j$ integers), with grid size $L
\times L$ and grid spacing $d x=d y$ and with periodic boundary
conditions. The Laplacians in Eq. (\ref{pfcp}) were approximated by
a simple discretization scheme
\begin{equation}
\frac{\partial^2}{\partial x^2}f_{i,j} = \frac{f_{i+1,j} + f_{i-1,j}
- 2 f_{i,j}}{(\Delta x)^2}.
\end{equation}
The standard Metropolis algorithm was used for simulations at a
given temperature. At each grid site $(i,j)$, we attempt to change
$\psi_{i,j}$ by an small amount $\Delta \psi$ with probability
$\min(1,e^{-\Delta H / k_BT} )$, using the Metropolis scheme,
where $\Delta H$ is the resulting change in the configurational
energy. Simulations were performed using, typically,  $L$ ranging
from $64$ to $ 224 $, grid spacing $dx= \pi/4$ and  $(4 - 2)
\times 10^6$ MC passes for equilibration and equal number of
passes for thermal averages. Because of the periodic boundary
conditions, the length of the system $L \ dx $ has to be a
multiple of the lattice constant of the pinning potential, $2 \pi
/k_s$. This condition imposes a restriction on the values of $k_s$
when calculations are done for fixed $L$ and $dx$.  Thus, to be
able to change $k_s$ continuously near a commensurate phase while
still keeping the same system size $L$, we choose a variable grid
spacing $d x = (k_{s}^{0}/k_{s})(\pi /4)$, where $k_{s}^{0}$ is
fixed such that $L \pi /4 = n (2 \pi /k_{s}^{0})$ ($n$ is an
integer). The parameters $r$ and $\mu$ were set to $r =-1/4$ and
$\mu =-0.1875$, corresponding to the hexagonal crystalline region
in the original PFC model without pinning \cite{elder02,elder04}.
The temperature $T$ is measured in units of $ F_0 k_B^{-1}$.

Near the phase transitions where long equilibration times are
required, we use the exchange MC method (parallel tempering)
\cite{nemoto,marinari}. This method is known to reduce significantly
the critical slowing down near phase transitions. Recently, it has
being used to study glassy incommensurate vortex lattices in 2D
superconducting arrays \cite{eg08}. In this method, many replicas of
the system with different temperatures $T_i$ in a range above and
below the critical temperature are simulated simultaneously and the
corresponding configurations are allowed to be exchanged with a
probability distribution satisfying detailed balance. The exchange
process allows the configurations of the system to explore the
temperature space, being cooled down and warmed up, and the system
can in principle escape more easily from metastable minima at lower
temperatures. Without the replica exchange step, the method reduces
to conventional MC simulations performed at different temperatures.
The method was implemented by performing MC simulations as described
above for each replica at different temperatures, simultaneously and
independently, for a few MC passes. Then exchange of pairs of
replica configurations at temperatures $T_i$ and $T_j$ and energies
$E_i$ and $E_j$ is attempted with probability
$\min(1,\exp(-\Delta))$, where $\Delta = (1/T_i - 1/T_j)(E_j -E_i)$,
using the Metropolis scheme.  We typically used $ 10^6$ MC passes
for equilibration with up to $30$ replicas and equal number of MC
passes for calculations of average quantities.

\section{Results and discussion}

\subsection{Ground-state phase diagram}

The ordered structures in the ground state were obtained by
simulations of the dissipative dynamical equations for the phase
field $\psi(\vec x)$, as in the previous work \cite{achim06}. The
reason for using this method instead of the MC method described in
Sec. III is that, at very low temperatures, MC simulations turned
out to be computationally less efficient due to the low acceptance
rate for MC moves. The dissipative dynamics should evolve the system
to the lower-energy state for arbitrary initial conditions.
Therefore the determination of the final configurations is
equivalent to finding the ground state for the effective Hamiltonian
in Eq. (\ref{part}), since $H[\psi]\equiv F[\psi]$. For the
free-energy functional of Eq. (\ref{pfcp}) with non-conserved field
the dynamical equations are
\begin{equation}
\frac{\delta \psi}{\delta t} = -\frac{\delta F}{\delta \psi}.
\label{deq}
\end{equation}
The ground-state phase diagram was obtained by finding the final
configurations where $\delta \psi/\delta t = 0$ starting from
different initial configurations, for different misfits $\delta_m$
and amplitudes of the pinning potential $V_0$. Typically, we start
with a hexagonal structure which is the minimum energy state in the
absence of the periodic potential ($V_0=0$) and then increase slowly
$V_0$ to the its final value for fixed $\delta_{\rm m}$. The
calculations are then repeated for different initials conditions.
The ground state is the final configuration corresponding to the
lowest energy.

\begin{figure}
\includegraphics[ bb= 4cm 13cm  20cm   24cm, width=7.5 cm]{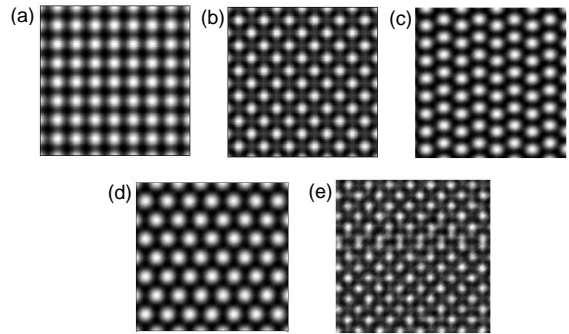}
\caption{Density plot of the phase field $\psi(\vec x)$ showing
commensurate (C) and incommensurate (IC) structures in the ground
state, depending on the amplitude of the pinning potential $V_0$ and
misfit parameter $\delta_{\rm m}$. (a) $(1 \times 1)$ C phase; (b)
$c(2 \times 2)$ C phase; (c) $(2 \times 1)$ C phase; (d) hexagonal
(full) IC phase and (e) IC phase with domain walls near the $c(2
\times 2)$ C phase. } \label{phases}
\end{figure}

Three low-order commensurate  structures were obtained as shown in
Fig. \ref{phases}: a $(1 \times 1)$ commensurate (C) phase, where
the peaks of $\psi(\vec x)$ coincide with the pinning potential
minima; a $c(2 \times 2)$ C phase, where they form a superstructure
with periodicity twice that of the pinning potential along the
principal directions and a $c(2 \times 1)$ phase  where the
superstructure has lattice periodicity  twice the pinning potential
along one of the directions. In addition, there are incommensurate
(IC) phases, where $\psi(\vec x)$ forms a hexagonal periodic
structures incommensurate with the pinning potential as in Fig.
\ref{phases}(d) or a structure of domain walls separating
commensurate regions as in Fig. \ref{phases}(e).

\begin{figure}
\includegraphics[ bb= 2cm 8cm  18cm   22cm,width=8 cm]{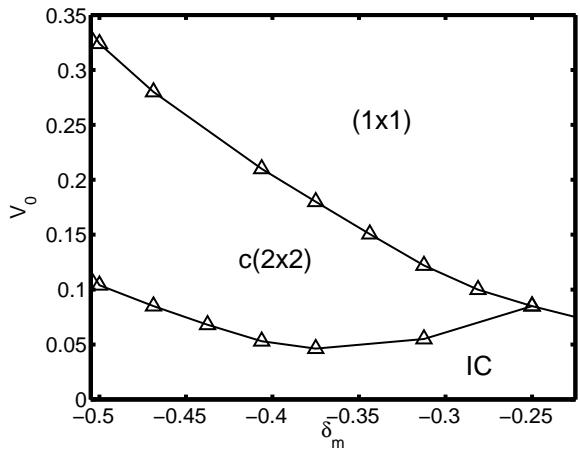}
\caption{Ground state phase diagram in terms of the pinning strength
$V_0$ and mismatch $\delta_{\rm m}$ for $-0.5 \le  \delta_{\rm m}
\le -0.225$} \label{diagT0a}
\end{figure}

\begin{figure}
\includegraphics[ bb= 2cm 8cm  18cm   22cm,width=8 cm]{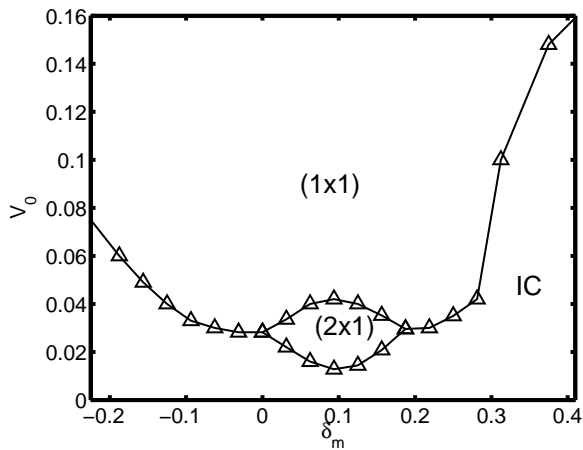}
\caption{Ground state phase diagram in terms of the pinning strength
$V_0$ and mismatch $\delta_{\rm m}$ for $-0.2 \le \delta_{\rm m} \le
0.4 $. } \label{diagT0b}
\end{figure}

The results of extensive numerical calculations using the dynamical
equation (\ref{deq}) are summarized in the phase diagram in Fig.
\ref{diagT0a} for $-0.5 \le \delta_{\rm m} \le -0.225$ and in Fig.
\ref{diagT0b} for $-0.2 \le \delta_{\rm m} \le 0.4  0$. The
transitions between the different phases were determined from the
change in the structure factor.

\subsection{Finite-temperature phase diagram}

We have studied the influence of thermal fluctuations and lattice
mismatch near the simplest commensurate structures, the $(1 \times
1)$ and $c(2 \times 2)$ commensurate phases, using the MC method
described in Sec. III. The phase diagrams were obtained by
monitoring the behavior of the structure factor and specific heat as
a function of temperature and lattice misfit. First, the structure
factor $S(\vec k)$ was calculated from the positions $\vec R_j$  of
the local peaks in the field $\psi(x,y)$ as
\begin{equation}
S(\vec k) = \langle \sum_{j,j'=1}^{N_P}\frac{1}{N_P}e^{-i\vec k
\cdot (\vec R_j -\vec R_{j'})} \rangle , \label{sf}
\end{equation}
where $N_p$ is the number of peaks.  To locate the peak positions
$\vec R_j$ for each configuration of $\psi(x,y)$ we have implemented
a computer algorithm \cite{ramosb} to find the local maxima in
$\psi(x,y)$ based on a particle location algorithm used in digital
image processing \cite{crocker}. Alternatively, a structure factor
can also be defined directly from the field $\psi(x,y)$ as
\begin{equation}
S_\psi(\vec k) = \frac{1}{L^2}\langle |\psi(\vec k)|^2 \rangle ,
\end{equation}
where $\psi(\vec k)$ is the Fourier transform of $\psi(\vec x)$.
While the two expressions give similar results the former expression
is better for characterizing a structural phase transition as it
includes only the ordering of the lattice and does not
simultaneously include fluctuations in the amplitude of the $\psi$
as the latter expression does.
The results described here were obtained using the definition in Eq.
(\ref{sf}). Second, the specific heat $c$ was calculated directly
from the average energy as $c= (1/L^2)(d \langle H \rangle/dT)$ and
from the fluctuation relation
\begin{equation}
c = \frac{1}{L^2 T^2}(\langle H^2 \rangle - \langle H \rangle^2).
\end{equation}
We checked that these two different ways of calculating $c$ gave the
same results, which indicates that the system is properly
equilibrated at a given temperature.

\begin{figure}
\includegraphics[ bb= 3cm 4cm  19cm   27cm, width=7.5 cm]{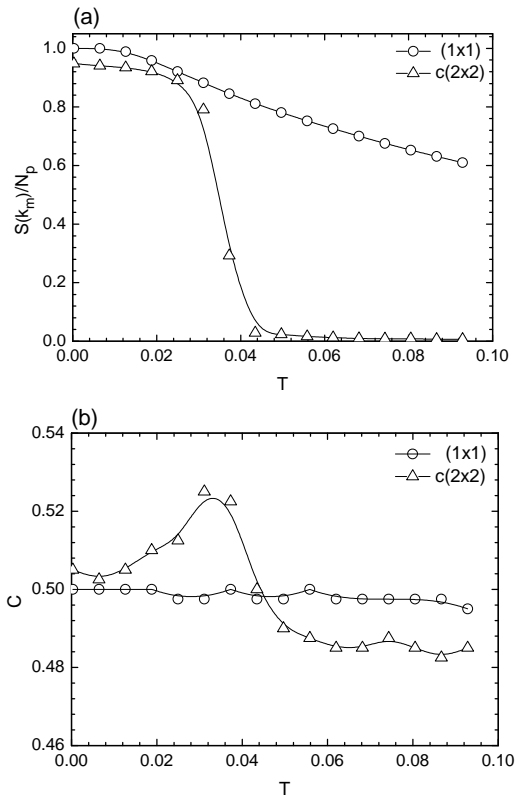}
\caption{ Temperature dependence of the scaled structure-factor peak
$S(k_m)/N_p$ (a) and specific heat $c$ (b) for the $(1 \times 1)$
commnesurate phase ( $\delta_{\rm m}=0$, $V_0=0.10$ ) and  $c(2
\times 2)$ commensurate phase ( $\delta_{\rm m}=-0.5$, $V_0=0.275$).
Here $k_m$ is the wave vector of the corresponding ordered
structure. }\label{sfheat}
\end{figure}

Figs. \ref{sfheat}a and \ref{sfheat}b  show the behavior of the
scaled structure-factor peak $S(k_m)/N_p$ and specific heat $c$ as a
function of temperature for the $(1 \times 1)$ and $c(2 \times 2)$
commensurate phases. Here $k_m$ is the magnitude of the wave-vector
corresponding to the local maximum in $S(\vec k)$. For the $(1\times
1)$ phase, $S(k_m)/N_p$ in Fig. \ref{sfheat}a decreases and broaden
with temperature but does not disappear at the highest temperature
where a liquid-like phase is expected. Also, there is no peak in the
corresponding specific heat in Fig. \ref{sfheat}b. This indicates
that the $(1\times 1)$ phase does not melt into a liquid-like phase
via a phase transition, instead there is a smooth crossover from a
low temperature highly ordered phase to a high temperature phase
where the pinning potential still induces some order in the peak
pattern of $\psi(\vec x)$. This behavior is expected on theoretical
grounds \cite{patry,haldane} since in this case the pinning
potential has the same symmetry as the commensurate phase and acts
as a constant external field on the displacement order parameter. On
the other hand, for the $c(2\times 2)$ commensurate phase,
$S(k_m)/N_p$ in Fig. \ref{sfheat}a drops sharply near the
temperature where there is a peak in the corresponding specific heat
in Fig. \ref{sfheat}b and this behavior is associated with the
melting of the ordered structure into a liquid-like phase.

The phase diagrams obtained near the $(1 \times 1)$ and $c(2 \times
2)$ commensurate phases as a function of misfit, strength of pinning
potential and temperature are shown in Figs. \ref{phd1x1} and
\ref{phdc2x2}, respectively. In addition to the melting transitions,
there are also commensurate-incommensurate transitions from the $c(2
\times 2)$ to IC phases and from $c(2 \times 2)$ to $(1 \times 1)$
phases which are identified by the change in the peak patterns in
the structure factor. A striking feature of the phase diagram is its
topology, which is strong dependent on the type of C phase. This
agrees qualitatively with theoretical predictions from simplified
models \cite{patry,haldane}. However, the present calculations were
not sufficiently accurate to determine the joining of the transition
lines near the incommensurate phases in Figs. \ref{phdc2x2}(a) and
(b). Hence, we were unable to investigate the theoretical
predictions \cite{haldane,rys} for the existence of an intermediate
liquid phase between the IC and C phases at finite temperature near
the  $c(2 \times 2)$ commensurate phase.

\begin{figure}
\includegraphics[ bb= 4cm 10cm  16cm   26cm,width=7.5 cm]{diag_(1x1)4.eps}
\caption{Phase diagrams near the $(1 \times 1)$ commensurate phase.
(a) at fixed pinning strength $V_0$ and (b) at fixed misfit
$\delta_{\rm m}$.}\label{phd1x1}
\end{figure}

\begin{figure}
\includegraphics[  bb= 4cm 10cm  16cm   26cm, width=7.5 cm]{diag_c(2x2)4.eps}
\caption{ Phase diagrams near the $c(2 \times 2)$ commensurate
phase. (a) at fixed pinning strength $V_0$ and (b) at fixed misfit
$\delta_{\rm m}$.}\label{phdc2x2}
\end{figure}

In addition to the topology of the phase diagram, the nature of
the phase transitions between the different phases is of
particular interest. We have investigated in detail the critical
behavior for the simplest case, corresponding to the melting
transition of the $c(2 \times 2)$ commensurate phase as a function
of the temperature. From symmetry considerations and Landau
free-energy expansions \cite{schick}, one expects that the
critical behavior  should be in the Ising universality class. To
verify if the phase field crystal model provides a correct
description of this behavior, we have performed a finite-size
scaling analysis of the melting of the $c(2 \times 2)$ phase as a
function of temperature, for a fixed value of misfit $\delta_{\rm
m}=-0.5$. Calculations were performed for increasing system sizes
for the specific heat, structure factor and a suitable
dimensionless Binder ratio \cite{binder}. We define the Binder
ratio $U_L(T)$ as
\begin{equation}
U_L = 2 - \frac{\langle |\rho(k_m)|^4 \rangle}{\langle |\rho(k_m)|^2
\rangle^2},
\end{equation}
where the order parameter $\rho(k_m)$ is the Fourier transform of
the of density of local peaks in the field $\psi(x,y)$ at $\vec
R_j$
\begin{equation}
\rho(k_m) = \frac{1}{N_P} \sum_{j=1}^{N_P} e^{-i \vec k_m \cdot \vec
R_j},
\end{equation}
evaluated at the wave vector $k_m$ corresponding to the local
maximum of the structure factor $S(k)$. The finite-size behavior of
$U_L$ provides an accurate determination of the critical temperature
$T_c$ and an estimate of the thermal critical exponent $\nu$ which
characterizes the divergent correlation length, $ \xi \propto
|T-T_c|^{-\nu}$, near the transition \cite{binder,binder97}. In the
high temperature disordered phase, the real and imaginary parts of
$\rho(k_m)$ fluctuate with a Gaussian distribution leading to $U_L
\rightarrow 0$ while at low temperature there is long-range order
with $\langle \rho(k_m) \rangle \ne 0$ and therefore $U_L
\rightarrow 1$ for $L \rightarrow \infty$. At the critical
temperature $T_c$, the ratio $U_L$ becomes independent of the system
size $L$ and therefore plots of $U_L(T)$ as a function of
temperature for different systems sizes should cross at the same
point, corresponding to the critical temperature $T_c$  of the
system in the thermodynamic limit. In the scaling regime
sufficiently close to $T_c$, the dimensionless $U_L(T)$ should
satisfy the scaling form
\begin{equation}
U_L(T)=\bar U ((T-T_c) L^{1/\nu}),
\end{equation}
where $\bar U(x)$ is a scaling function. Since the slope ($\partial
U_L(T)/\partial T$) evaluated at $T_c$ is proportional to
$L^{1/\nu}$, an estimate of $\nu$ can be obtained \cite{binder97}
from a log-log plot of this quantity against $L$.

\begin{figure}
\includegraphics[  bb= 2cm 2cm  20cm   26cm, width=7.5 cm]{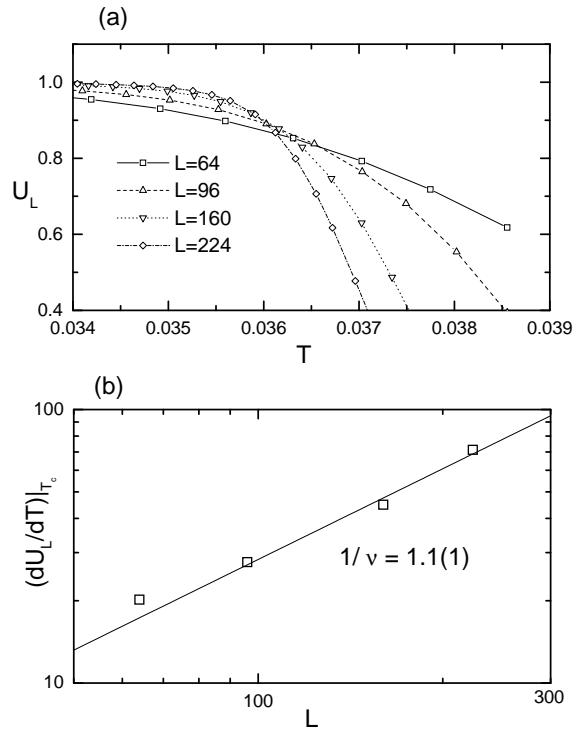}
\caption{ (a) Temperature dependence of the Binder ratio $U_L(T)$
for different system sizes $L$, near the melting transition of the
$c(2 \times 2)$ commensurate phase at $\delta_{\rm m}=-0.5$ and
$V_0=0.275$; (b) Estimate of the thermal critical exponent $\nu$
from the log-log plot of $\frac{\partial}{\partial T} U_L(T)$ at
$T_c$ against $L$ for the three largest system sizes.} \label{UxT}
\end{figure}

Fig. \ref{UxT}(a) shows the temperature behavior of $U_L$ for
different system sizes. Although the curves do not cross precisely
at the same point, for the three largest system sizes the curves
intersect approximately at $T_c \approx 0.03599(20)$. The lack of
intersection at a common point should be due to statistical errors
and corrections to scaling.  Fig. \ref{UxT}b shows a log-log of
$\delta U_L(T)/\delta T$ evaluated at the estimated $T_c$ against
$L$ from where we obtain $1/\nu =1.1(1)$. This estimate of $\nu$ is
in agreement with the exact value for the thermal exponent of the
two-dimensional Ising model, $\nu =1$.

\begin{figure}
\includegraphics[  bb= 2cm 1cm  20cm   26cm, width=7.5 cm]{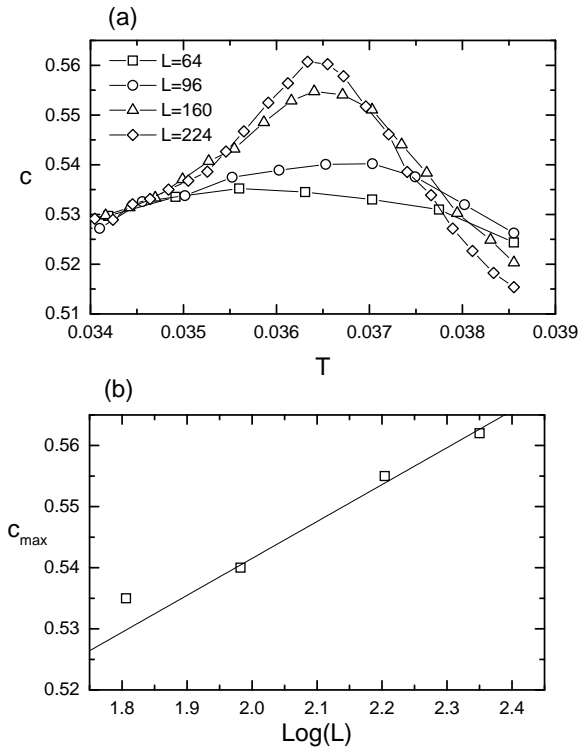}
\caption{ (a) Temperature dependence of the specific heat $c$ for
different system sizes $L$, near the melting transition of the $c(2
\times 2)$ commensurate phase at $\delta_{\rm m}=-0.5$ and
$V_0=0.275$; (b) Specific heat maxima $c_{max}$ in a linear-log plot
indicating a logarithmic behavior as a function of $L$ for the three
largest systems sizes.} \label{CxT}
\end{figure}

The finite-size behavior of the specific heat $c$ is also consistent
with the Ising universality class, where the specific heat exponent
is $\alpha =0$. Fig. \ref{CxT}(a) shows the temperature behavior of
the specific heat for the different systems sizes. From finite-size
scaling, the maximum of $c$ should scale as $L^{\alpha/\nu}$, which
corresponds to a logarithmic behavior $C_{max} \propto \log L$, if
$\alpha = 0$. Fig.  \ref{CxT}(b) shows that the linear-log plot of
the specific heat maximum $c_{max}$ against $L$ is indeed consistent
with a logarithmic behavior for the three largest system sizes.

\begin{figure}
\includegraphics[  bb= 2cm 2cm  20cm   18cm, width=7.5 cm]{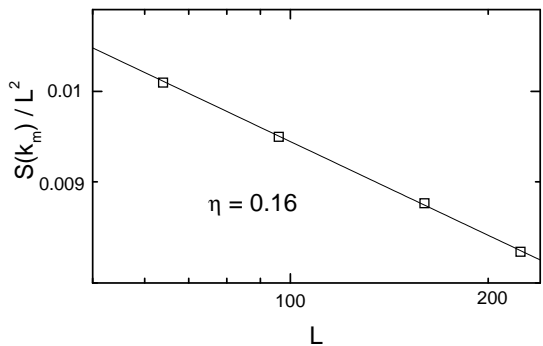}
\caption{ Estimate of the correlation function exponent  $\eta $
from the log-log plot of $S(k_m)/L^2 $ at $T_c$ against $L$ for the
three largest systems sizes.} \label{SxL}
\end{figure}

The correlation function exponent $\eta$ can also be estimated from
the expected finite-size behavior of the structure factor, which
should scale as $S(k_m) \propto L^{2-\eta}$ at $T_c$. Fig. \ref{SxL}
shows a log-log plot of $S(k_m)/L^2$ evaluated at the above
estimated $T_c$ against $L$ from where we find $\eta = 0.15$. Unlike
the above estimates of $\nu$ and $\alpha$, this estimate of $\eta$
is much smaller than the known exact value for the Ising model,
$\eta =0.25$. This large discrepancy could be due to corrections to
scaling. However, to take such corrections into account would
require more accurate data and larger system sizes, which is beyond
the scope of the present work.

\section{Conclusions}

In the present work we have studied the effects of thermal
fluctuations in the phase field crystal model with an external
pinning potential \cite{achim06} using a non-conserved version of
the model and Monte Carlo simulations. We have determined  the phase
diagram as a function of temperature and mismatch near low-order
rational commensurate phases. The results show a rich phase diagram
with commensurate, incommensurate and liquid-like phases with a
topology dependent on the type of ordered structure qualitatively
consistent with predictions of simplified models of pinned lattice
systems \cite{patry,haldane}. In particular, we find that the
melting transition for the $c(2 \times 2)$ commensurate phase is
consistent with the Ising universality class, which is expected from
analytical arguments based on symmetry considerations and Landau
free-energy expansion \cite{schick}. Our results demonstrate that
the PFC model and the MC method employed here can be used to study
specific lattice systems by adjusting the parameters of the model to
match the experimental structure factor \cite{elder04,elder07} and
choosing a suitable pinning potential.

\begin{acknowledgments}

J.A.P.R. acknowledges the support from Secretaria da Administra\c
c\~ao do Estado da Bahia. E.G. was supported by Funda\c c\~ao de
Amparo \`a Pesquisa do Estado de S\~ao Paulo - FAPESP (Grant no.
07/08492-9). C.V.A. acknowledges the support from Magnus Ehrnrooth
(Finland) for providing a travel grant. K.R.E. acknowledges the
support from NSF under Grant No. DMR-0413062. This work has also
been supported by joint funding under EU STREP 016447 MagDot and NSF
DMR Award No. 0502737, and by the Academy of Finland through its
Center of Excellence COMP grant. We also acknowledge CSC-Center for
Scientific Computing Ltd. for allocation of computational resources.

\end{acknowledgments}


\begin{thebibliography}{00}
\bibitem{foot1} The wave-vector of the periodic lattice is strictly
only one in a one-mode approximation and may deviate by a few
percent when higher order modes are considered.
\end{thebibliography}
\end{document}